\documentclass[]{spie}
 \usepackage[american]{babel}
\usepackage{amsmath,amsfonts,amssymb}
\usepackage{graphicx}
\usepackage[colorlinks=true, allcolors=blue]{hyperref}
\usepackage{subfigure}
\usepackage{multirow}
\usepackage{float}
\usepackage{mathabx}
\usepackage{epstopdf}
\usepackage{booktabs}
\usepackage{cleveref}

\title{End-to-end simulations of a near-infrared pyramid sensor on Keck II}

\author[a]{C. Plantet}
\author[a]{G. Agapito}
\author[a]{C. Giordano}
\author[a]{S. Esposito}
\author[b]{P. Wizinowich}
\author[c]{C. Bond}
\affil[a]{INAF - Osservatorio di Arcetri, 50125 Firenze, Italy}
\affil[b]{W. M. Keck Observatory, 65-1120 Mamalahoa Hwy., Kamuela, HI 96743, USA}
\affil[c]{Institute for Astronomy, University of Hawaii, 640 N. Aohoku Place, Hilo, HI 96720}

\authorinfo{Further author information: \\C.P.:
  E-mail: plantet@arcetri.astro.it}

\begin{document} 
\maketitle

\begin{abstract}
The future upgrade of Keck II telescope’s adaptive optics system will include a pyramid wavefront sensor working in the near-infrared (J and H band). It will benefit from the recently developed avalanche photodiode arrays, specifically the SAPHIRA (Selex) array, which provides a low noise ($<$ 1 e- at high frame rates). The system will either work with a natural guide star (NGS) in a single conjugated adaptive optics system, or in a laser guide star (LGS) mode. In this case, the pyramid would be used as a low-order sensor only. We report on a study of the pyramid sensor's performance via end-to-end simulations, applied to Keck's specific case. We present the expected Strehl ratio with optimized configurations in NGS mode, and the expected residual on low orders in LGS mode. In the latter case, we also compare the pyramid to LIFT, a focal-plane sensor, demonstrating the ability of LIFT to provide a gain of about 2 magnitudes for low-order sensing.
\end{abstract}


\section{INTRODUCTION}

The future upgrade of Keck II telescope's Adaptive Optics (AO) system \cite{wizinowich1} will include a pyramid wavefront sensor \cite{ragazzoni} working in the near-infrared (J and H band)\cite{wizinowich3}. The main goal of this upgrade is to perform direct imaging and slit spectroscopy of exoplanets around M dwarfs. The flux from these stars is very faint at optical wavelengths, but sufficient in the near-infrared to use as NGSs in a single conjugated AO system, given the adequate detector technology. The recently developed avalanche photodiode arrays, such as the SAPHIRA (Selex), provide a low noise ($<$ 1 e- at high frame rates) and are thus suitable for this application \cite{feautrier}. In addition to the NGS mode, the system will also provide a LGS mode. In this case, the pyramid would be used as a low-order sensor only. We report on a study of the pyramid sensor's performance via end-to-end simulations made with PASSATA \cite{agapito}. After a quick summary of the simulation parameters (section \ref{params_sim}), we present the expected Strehl ratio in NGS mode (section \ref{ngs_mode}), and the expected residual on low orders in LGS mode (section \ref{lgs_mode}). In the latter case, the pyramid will not benefit from a hardware rebinning of pixels, and thus will not be in a fully optimized configuration. For this reason, we also compare the pyramid to LIFT \cite{meimon}, a focal-plane sensor, that could provide a better low-order estimation at low flux.

\section{Simulations parameters}
\label{params_sim}

We list in Table \ref{params} the simulation parameters used for the different cases of this study. The chosen values for the parameters that are optimized (modulation, frequency, control gain\dots) are given in each specific case, and we only state here the explored ranges of values. The wavefront modes (turbulent Karhunen-Lo\`eve and Zernike) are considered to be perfectly reproduced by the deformable mirror (DM). In all cases, the correction is made with an integrator command, and the delay depends on the frequency, with the following rules (taken from ERIS simulations experience \cite{quiros}):
\begin{itemize}
\item $f > 666\ Hz$ : 3 frames delay
\item $333\ Hz < f \leq 666\ Hz$ : 2 frames delay
\item $f \leq 333\ Hz$ : 1 frame delay
\end{itemize}

\begin{table}[H]
\caption{Simulation parameters.} 
\label{params}
\begin{center}       
\begin{tabular}{|c|c|c|c|c|c|c|} 
\hline
\multirow{2}{*}{Parameter} & \multicolumn{2}{c|}{NGS mode} & \multicolumn{2}{c|}{LGS mode} \\ \cline{2-5}
 & 20$\times$20 & 32$\times$32 & 20$\times$20 & 32$\times$32 \\
 \hline
Sensing band & \multicolumn{4}{c|}{1.5 $\mu m$ - 1.8 $\mu m$ (H band)} \\
\hline
Pupil mask & \multicolumn{2}{c|}{Keck primary on 512 pixels} & \multicolumn{2}{c|}{Keck primary on 256 pixels} \\ 
\hline
Mode basis & 250 KL modes & \multicolumn{3}{c|}{245 KL modes + 5 first Zernike} \\
\hline
Total transmission (including QE) & \multicolumn{4}{c|}{0.3} \\
\hline
Sky background in H & \multicolumn{4}{c|}{14 mag/arcsec$^2$} \\
\hline
Seeing & \multicolumn{4}{c|}{0.63"} \\
\hline
 Layers' altitudes (km) &  \multicolumn{4}{c|}{0, 500, 1000, 2000, 4000, 8000, 16000}  \\
\hline
  $C_n^2$ profile (normalized in energy) &  \multicolumn{4}{c|}{0.517, 0.119, 0.063, 0.061, 0.105, 0.081, 0.054}  \\
\hline 
  Mean wind speed & \multicolumn{4}{c|}{9.5 m/s}  \\
\hline
  Zenith angle & \multicolumn{4}{c|}{30$^\degree$}  \\
\hline
  Subaperture size & 0.5625 m & 0.35 m & 0.5625 m & 0.35 m  \\
\hline
APD gain &   \multicolumn{4}{c|}{30} \\
\hline
Excess noise factor & \multicolumn{4}{c|}{1.4} \\
\hline
Read-out noise & 0.1 or 1 e$^-$ & 1 e$^-$  & 0.8 e$^-$ & 1 e$^-$ \\
\hline
Dark current & 0 or 100 e$^-$/s & 20 e$^-$/s & 100 e$^-$/s & 20 e$^-$/s \\
\hline
  Frequency range &  \multicolumn{2}{c|}{300-1000 Hz} & 200-1000 Hz &   \\
\hline
  \multirow{2}{*}{Control gain range} & \multicolumn{2}{c|}{\multirow{2}{*}{0.1-0.6}} & LIFT: 0.1-0.6 & LIFT: 0.1-0.6  \\
  & \multicolumn{2}{c|}{} & Pyramid: 0.15-10 & Pyramid:0.25-5 \\
\hline
  Pyramid modulation radius range & 1-3 $\lambda/D$ & 1-2 $\lambda/D$ & \multicolumn{2}{c|}{0-2 $\lambda/D$} \\
\hline
  FoV & \multicolumn{4}{c|}{1"} \\
\hline
  Additional HO residual (non corrected) & \multicolumn{2}{c|}{60 nm} & 0 nm & 60 nm \\
\hline
\end{tabular}
\end{center}
\end{table}

To these parameters, we add the following precisions for the LGS mode:
\begin{itemize}
\item High-order loop parameters:
\begin{itemize}
\item Sensor: SH 20$\times$20 with quad-cells estimating 250 modes.
\item LGS = high flux point source at finite distance.
\item Tip/tilt filtered and replaced by a residual jitter of 106 mas rms + turbulent tip/tilt.
\item Control gain: 0.3.
\end{itemize}
\item Focus loop (only in 32$\times$32 case):
\begin{itemize}
\item Correction frequency: 10 Hz.
\item Input: focus residual from high-order control + sinusoid of period 5 seconds and amplitude 100 nm (80 nm rms).
\item Control gain range: 0.1-1 for LIFT, 0.1-4 for the pyramid.
\end{itemize}
\end{itemize}

Finally, for consistency with the error budget used in a previous study \cite{wizinowich3}, we add a constant error of 165 nm rms to the residual in NGS mode, representing miscellaneous errors from undetermined sources.

\section{NGS mode}
\label{ngs_mode}
In this section, we study the performance of the pyramid, in terms of Strehl ratio, for different pupil samplings. We first considered a pupil sampling of 20$\times$20 subapertures, in agreement with the current DM's number of actuators. However, the DM should be upgraded to a MEMS 32$\times$32. We thus study in a second step the impact of increasing the pupil sampling to 32$\times$32, or to 40$\times$40 for robustness reasons.

\subsection{Pyramid 20$\times$20}
\label{ngs_mode_20}

\begin{figure}[H]
\centering
\subfigure[Low noise]{
\includegraphics[width=.3\linewidth]{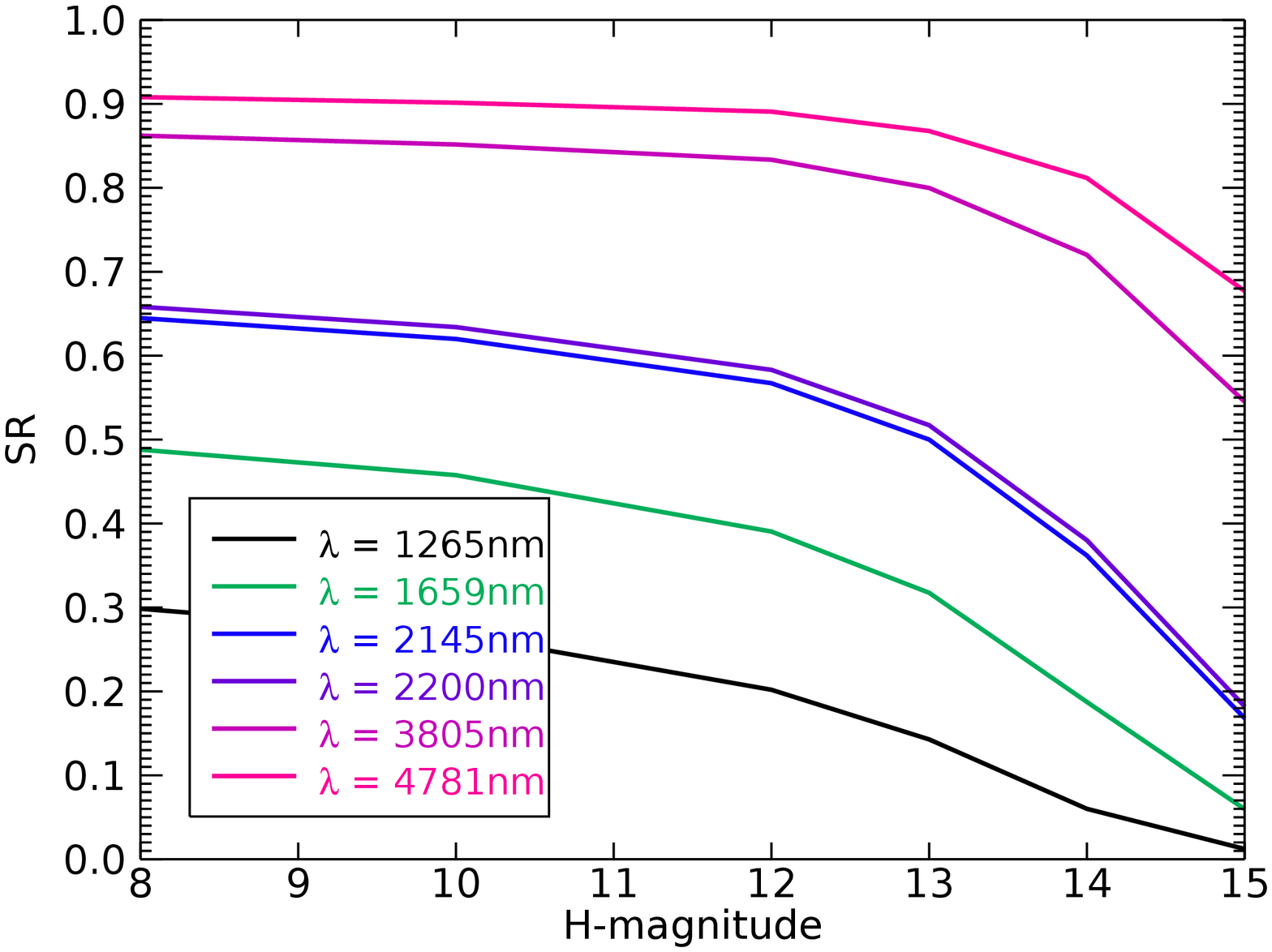}
\label{pyr_ngs_low_noise}
}
\subfigure[High noise]{
\includegraphics[width=.3\linewidth]{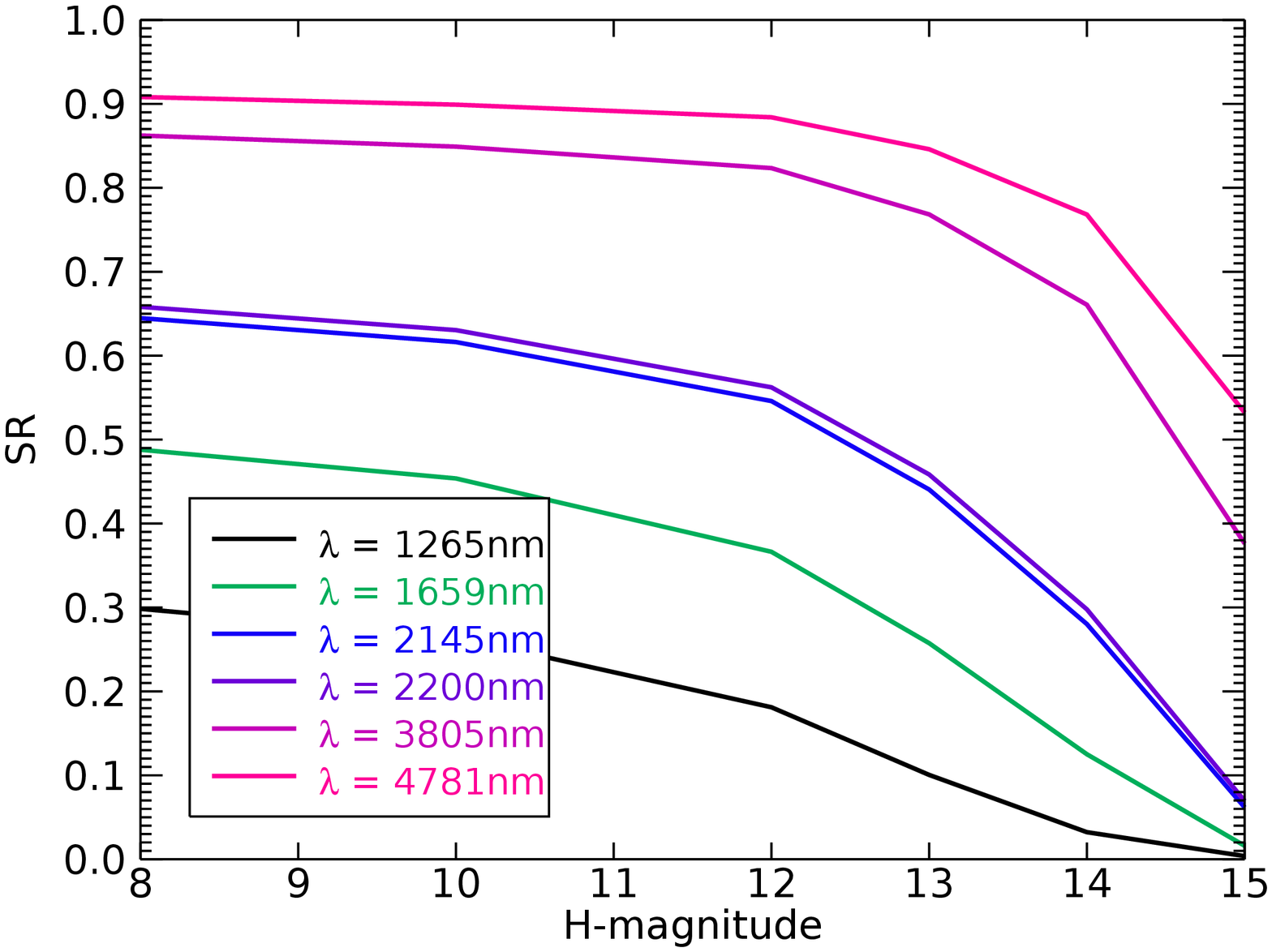}
\label{pyr_ngs_high_noise}
}
\subfigure[Comparison]{
\includegraphics[width=.3\linewidth]{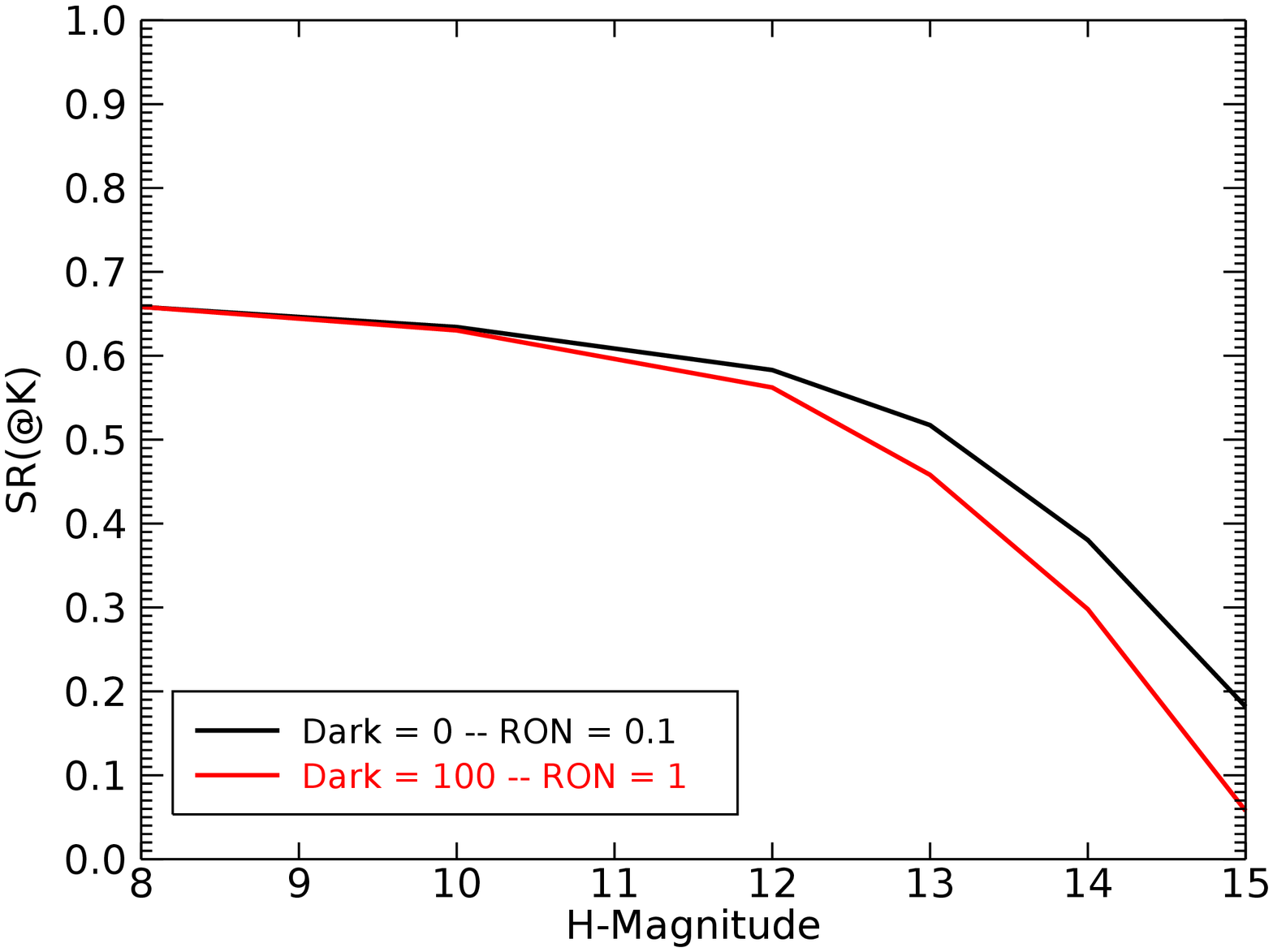}
\label{pyr_ngs_low_vs_high_noise}
}
\label{pyr_ngs}
\caption{Strehl ratio as a function of H magnitude with a pyramid 20$\times$20 in NGS mode. Left: Low noise case. Middle: High noise case. Right: Comparison of low and high noise cases in K band.}
\end{figure}

We present here the results of the simulations with a pyramid 20$\times$20 in NGS mode (Fig. \ref{pyr_ngs}). The parameters, listed in Table \ref{params_ngs20},  are optimized in the ranges described previously to get the highest Strehl ratio. This optimization is simply made by running simulations going through the whole set of parameters and selecting the best one.

As we lacked information on the detector's noise, we considered two cases: low noise (no dark current, read-out noise = 0.1 e$^-$) and high noise (dark current = 100 e$^-$/s, read-out noise = 1 e$^-$). The difference between those two cases is not very significant (0.5 magnitude at faint end).

\begin{table}[H]
\caption{Optimized parameters (high noise/low noise) for the pyramid 20$\times$20 in NGS mode.} 
\label{params_ngs20}
\begin{center}       
\begin{tabular}{|c|c|c|c|c|c|c|} 
\hline
Magnitude & 8 & 10 & 12 & 13 & 14 & 15 \\
\hline
Frequency (Hz) & 1000/1000 & 1000/1000 & 1000/600 & 1000/500 & 600/300 & 600/300 \\ 
\hline
Number of modes & 250/250 &  250/250 & 170/152 & 135/104 & 65/54 & 44/14 \\
\hline
Gain & 0.3/0.3 & 0.2/0.2 & 0.15/0.25 & 0.15/0.3 & 0.3/0.55 & 0.3/0.6 \\
\hline
Modulation radius ($\lambda/D$) & 1.5 & 1.5 & 1.5 & 1.5 & 1.5 & 2  \\
\hline
\end{tabular}
\end{center}
\end{table}

Overall, what we observe on the optimization of the parameters when we are going towards lower fluxes is: 
\begin{itemize}
\item Decrease in frequency: to collect more flux and reduce the noise error.
\item Increase the gain: we cannot remove the background in pyramid images, it is thus taken into account in the normalization when computing the slopes. In the end, the slopes are proportional to the ratio star flux/background, which decreases with respect to the magnitude. A higher gain is needed to compensate that effect. The increase in gain is also needed at lower frequencies, as the correction is done less often.
\item Increase in modulation: at low flux, the noise error makes the pyramid work in non-linear regime. The modulation reduces the non-linearity error, at the price of a lower sensitivity (hence greater noise error). A trade-off is made between those two errors to reach the lowest overall error. At high flux, using a high modulation lowers the non-linearity error.
\item Decrease the number of modes: estimating less modes improves the noise propagation behavior at low orders.
\end{itemize}

These results are consistent with the ones presented in an earlier study\cite{wizinowich3}, with a difference of only a few percents of Strehl ratio in K band.

\subsection{Impact of a finer pupil sampling}
Having a finer pupil sampling allows us to estimate more modes at high flux, but lowers the signal-to-noise ratio (SNR) at low flux. We consider here only the impact at low flux, as it corresponds to more practical cases and is more critical for the system design.

We simulated two different pupil samplings: 32$\times$32 and 40$\times$40 (in that case, only the subaperture size from Table \ref{params} is changed). The first one matches the MEMS mirror sampling, while the second would help calibrate misregistration errors and thus gain in robustness.

The performance and optimized parameters at magnitude 14 are given in Table \ref{params_ngs32}, for a dark current of 20 e$^-$/s and a read-out noise of 1 e$^-$. The performance for the high noise case of the pyramid 20$\times$20 is recalled for reference. It should be noted that the dark current does not have a significant impact here, the results can thus be fairly compared.

\begin{table}[H]
\caption{Optimized parameters and Strehl ratios for the pyramid 32$\times$32 and 40$\times$40 in NGS mode.} 
\label{params_ngs32}
\begin{center}       
\begin{tabular}{|c|c|c|c|c|c|} 
\hline
 & Frequency (Hz) & Number of modes & Gain & Modulation radius ($\lambda/D$) & Strehl ratio (K)  \\
 \hline
20$\times$20 & 600 & 65 & 0.3 & 1.5 & 29.8\% \\ 
\hline
32$\times$32 & 200 & 65 & 0.75 & 1.5 & 26.7\% \\ 
\hline
40$\times$40 & 200 & 65 & 0.75 & 1.5 & 25.1\% \\
\hline
\end{tabular}
\end{center}
\end{table}

The finer pupil sampling does not have a strong impact on performance: the loss of Strehl in Ks is 3\% for the 32$\times$32 and 5\% for the 40$\times$40. Hence, it seems a reasonable choice to go towards a 40$\times$40 sampling, making the system more reliable without a significant loss of performance at low flux.

\section{LGS mode}
\label{lgs_mode}
The goal of this section is to assess the achievable residual on tip/tilt and focus in LGS mode, for a NGS on axis or at 15" off axis. We compare the pyramid to LIFT, in order to evaluate the gain of having a focal-plane sensor for this low-order estimation. Indeed, as we cannot do a hardware rebin of pixels on the camera, the pyramid would still utilise a fine sampling and would thus have poorer noise propagation properties for low-order estimation than with a coarse sampling. 

As in the previous section, we first considered a pyramid with 20$\times$20, and then checked the impact of a finer sampling. For LIFT, the only design parameter that will have an impact on the performance is the pixel scale. We consider here a pixel of 15 or 30 mas, corresponding respectively to a Nyquist and a Nyquist/2 sampling in H band. 

\subsection{Pyramid 20$\times$20}

In this part, we only evaluate the residual on tip/tilt, as it is the most important feature of the low-order sensor. The focus estimation will be included in the next section. For practical reasons, the number of reconstructed modes for the pyramid is either 2 (lowest noise error) or 250 (lowest aliasing error).

\begin{figure}[H]
\centering
\subfigure[On axis]{
\includegraphics[width=.4\linewidth]{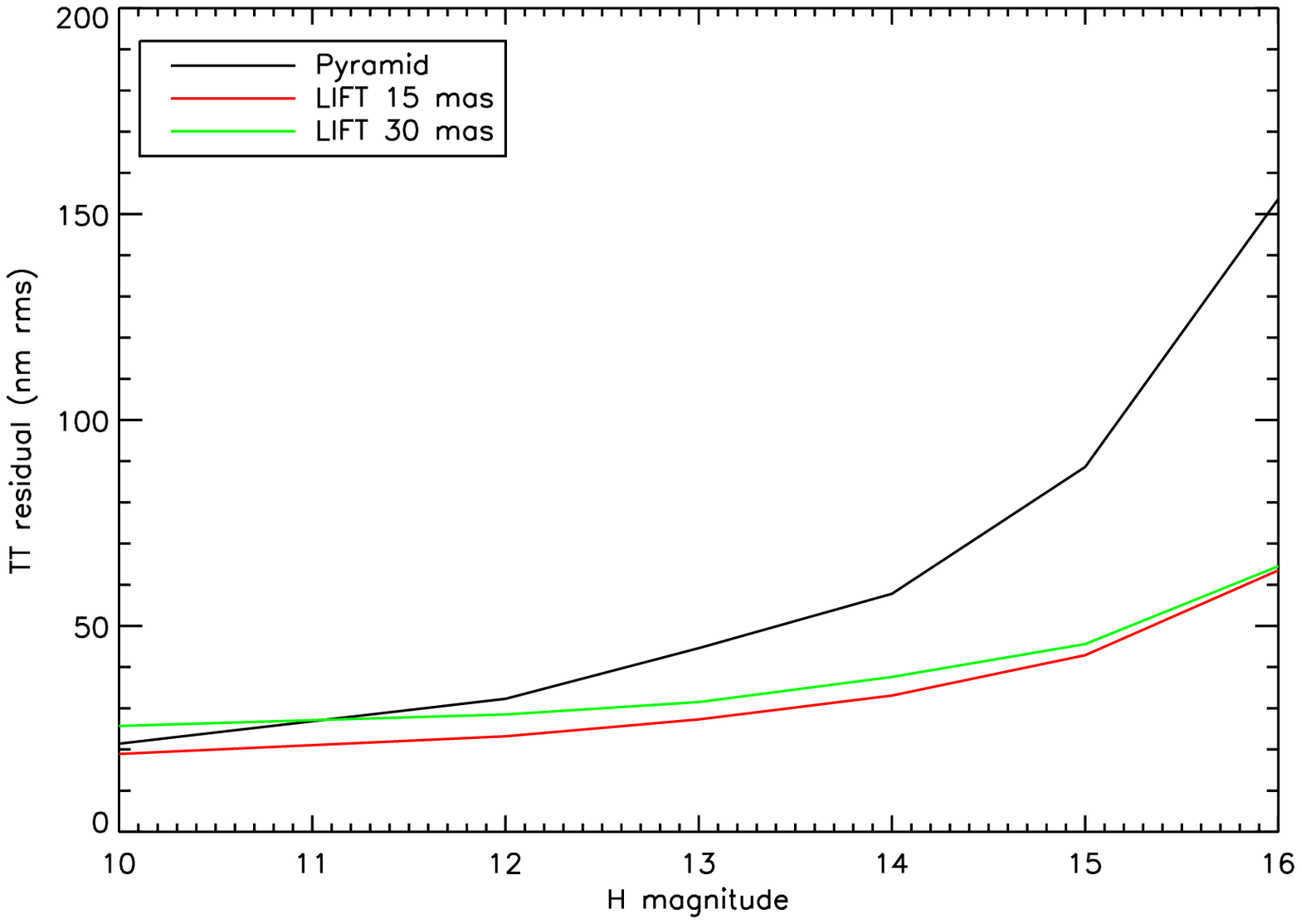}
\label{comp_lift_pyr_onaxis}
}
\subfigure[15" off axis]{
\includegraphics[width=.4\linewidth]{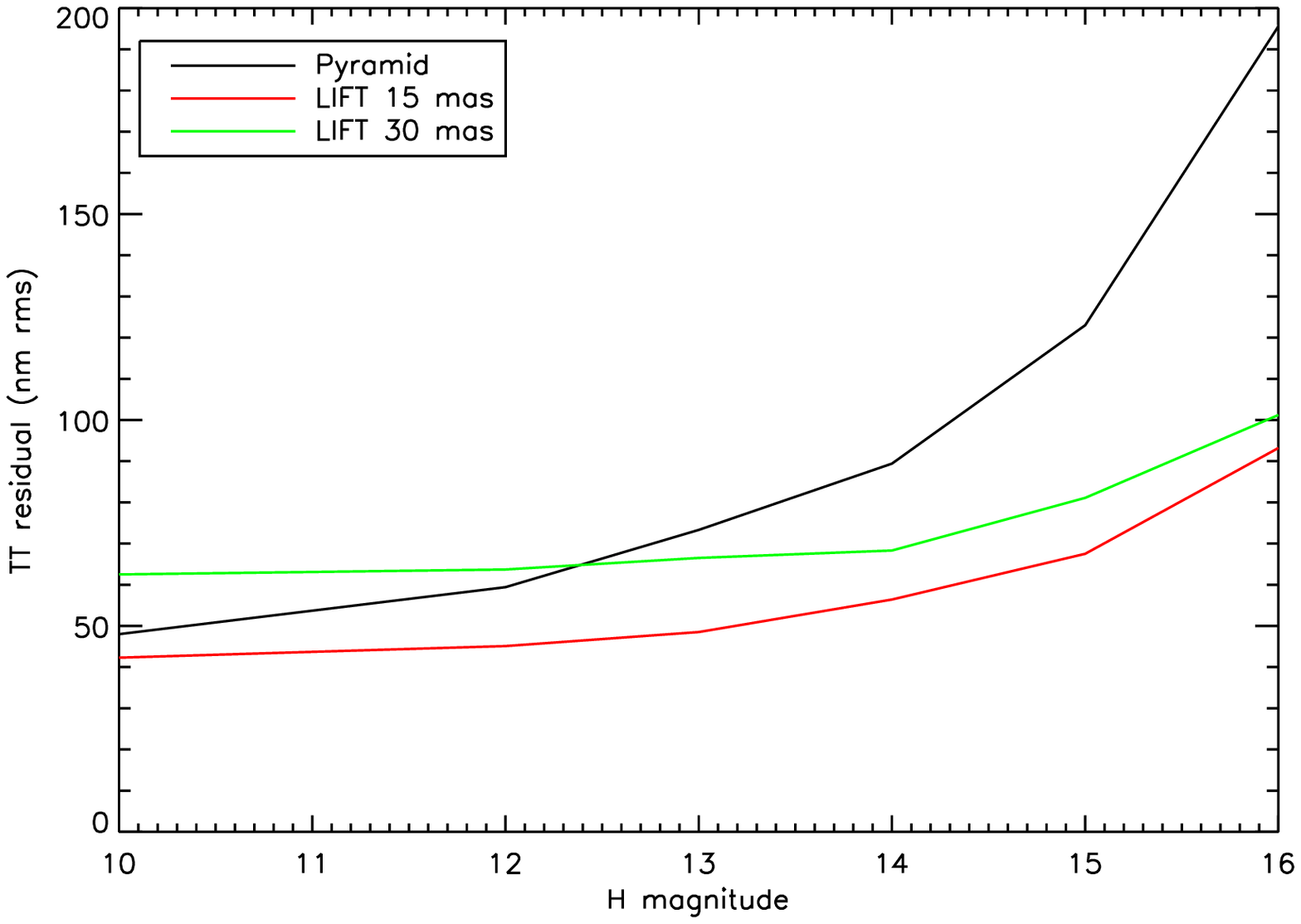}
\label{comp_lift_pyr_15asec}
}
\label{comp_lift_pyr}
\caption{Comparison of the tip/tilt residual obtained with LIFT or the pyramid 20$\times$20 in LGS mode. Left: NGS on axis. Right: NGS at 15" off axis.}
\end{figure}

We list in \Cref{params_lgs_pyr20,params_lgs_lift15,params_lgs_lift30} the optimized parameters for LIFT and the pyramid in each case, as well as the residual on tip/tilt.  The residuals obtained with LIFT and the pyramid are compared in Fig. \ref{comp_lift_pyr}. We find that LIFT provides a gain of up to 2 magnitudes over the pyramid, either on axis or off axis. 

The behavior of the optimized parameters for the pyramid is as described in section \ref{ngs_mode_20}. In particular, we can notice an increase in modulation at high flux when going off-axis: this is due to the increase in amplitude of high-order modes, for which the linearity must be improved. The flux is sufficiently high in that case to use a strong modulation without a significant impact on noise error. 

As concerns LIFT's sampling, the pixel of 30 mas benefits from better noise propagation properties (better SNR/pixel), but does not provide any significant improvement of the performance. On the contrary, it is less efficient off axis, or at high flux in general. Indeed, the signal from high orders, normally far from the spot center, gets more easily mixed with the low orders signal, which is within the spot center. This aliasing error is visible at high flux, where the noise error is negligible, and gets higher when going off axis, where the Strehl ratio is lower. The overall aliasing + noise error is in the end always better with the 15 mas pixel for the considered magnitudes.

\begin{table}[H]
\caption{Optimized parameters (on axis/off axis) for the pyramid 20$\times$20 in LGS mode.} 
\label{params_lgs_pyr20}
\begin{center}       
\begin{tabular}{|c|c|c|c|c|c|c|} 
\hline
Magnitude & 10 & 12 & 13 & 14 & 15 & 16 \\
\hline
Frequency (Hz) & 1000/1000 & 1000/1000 & 1000/1000 & 1000/1000 & 200/200 & 200/200 \\ 
\hline
Number of modes & 250/250 &  250/250 & 2/250 & 2/2 & 2/2 & 2/2 \\
\hline
Gain & 0.5/0.25 & 0.5/0.65 & 0.75/0.65 & 1/1 & 3.5/3.5 & 5/5 \\
\hline
Modulation radius ($\lambda/D$) & 0/2 & 0/0 & 0/0 & 0/0 & 0/0 & 0/1  \\
\hline
TT residual (nm rms) & 21.4/48 & 32.3/59.4 & 44.6/73.3 & 57.8/89.4 & 88.6/123 & 153.7/195.6 \\
\hline
\end{tabular}
\end{center}
\end{table}

\begin{table}[H]
\caption{Optimized parameters (on axis/off axis) for LIFT with a 15 mas pixel in LGS mode.} 
\label{params_lgs_lift15}
\begin{center}       
\begin{tabular}{|c|c|c|c|c|c|c|} 
\hline
Magnitude & 10 & 12 & 13 & 14 & 15 & 16 \\
\hline
Frequency (Hz) & 1000/1000 & 1000/1000 & 1000/333 & 333/333 & 333/200 & 200/200 \\ 
\hline
Gain & 0.3/0.3 & 0.3/0.3 & 0.2/0.5 & 0.4/0.5 & 0.4/0.4 & 0.4/0.3 \\
\hline
TT residual (nm rms) & 18.9/42.3 & 23.2/45.1 & 27.3/48.5 & 33.1/56.4 & 42.9/67.5 & 63.5/93.2 \\
\hline
\end{tabular}
\end{center}
\end{table}

\begin{table}[H]
\caption{Optimized parameters (on axis/off axis) for LIFT with a 30 mas pixel in LGS mode.} 
\label{params_lgs_lift30}
\begin{center}       
\begin{tabular}{|c|c|c|c|c|c|c|} 
\hline
Magnitude & 10 & 12 & 13 & 14 & 15 & 16 \\
\hline
Frequency (Hz) & 1000/1000 & 1000/1000 & 1000/1000 & 333/1000 & 333/500 & 200/333 \\ 
\hline
Gain & 0.4/0.3 & 0.3/0.2 & 0.2/0.2 & 0.5/0.2 & 0.4/0.2 & 0.4/0.3 \\
\hline
TT residual (nm rms) & 25.7/62.5 & 28.5/63.7 & 31.5/66.5 & 37.6/68.3 & 45.6/81.1 & 64.5/101.2 \\
\hline
\end{tabular}
\end{center}
\end{table}

\subsection{Finer pupil sampling}
We now check the impact of having a finer-sampled pyramid on the performance at magnitude 14. We also verify that we have a correct estimation of focus on both sensors, and we study the possibility of having 10 mas pixels on LIFT (for design simplicity reasons).

The parameters and results are given in Table \ref{params_lgs32}, for an optimization on tip/tilt correction only (the focus is then added with the same loop parameters).

On axis, there is a clear advantage using LIFT for tip/tilt estimation, with a factor 2 in rms residual. The estimation of focus does not affect the estimation of tip/tilt, whatever the sensor, and LIFT has a slight advantage on this mode as well (factor 1.4).

Off axis, we still have a better estimation of tip/tilt with LIFT, but with less difference (factor 1.4 at best). LIFT's performance is actually weakly dependent on the frequency: at 1000 Hz, the tip/tilt residual is increased by approximately 5 nm for the 15 mas and 10 mas pixels. The focus correction is similar in all cases, and the tip/tilt estimation is always affected. This might be the result of the sensors getting less and less linear when lowering the Strehl ratio (aliasing error discussed in the previous section). The effect seems stronger on LIFT at fine samplings (15 and 10 mas), but for these cases, as well as the pyramid, it is equivalent to adding an error of approximately 30-35 nm rms. For LIFT with 30 mas pixels, this error is lower, around 20 nm rms. This might be due to the fact that the tip/tilt estimation alone was already affected by non-linear effects. 

\begin{table}[H]
\caption{Optimized parameters and residuals (on axis/off axis) for the pyramid 32$\times$32 and LIFT in LGS mode, with the estimation of focus.} 
\label{params_lgs32}
\begin{center}       
\begin{tabular}{|c|c|c|c|c|} 
\hline
 & LIFT 30 mas & LIFT 15 mas & LIFT 10 mas & Pyramid \\
\hline
Frequency (Hz) & 333/1000 & 333/200 & 333/333 & 333/333 \\ 
\hline
Modulation radius ($\lambda/D$) & \multicolumn{3}{c|}{}  & 0/0  \\
\hline
Gain & 0.4/0.1 & 0.4/0.5 & 0.4/0.3 & 4/4 \\
\hline
TT residual (nm rms) & 39.3/80.5 & 34.3/67.5 & 34.7/69.8 & 68.9/95.9 \\
\specialrule{.1em}{.05em}{.05em}
TT residual (with focus) & 36.2/82.7 & 33.6/75.9 & 34.8/77.7 & 69.3/100 \\
\hline
Gain on focus & 0.6/0.4 & 0.5/0.4 & 0.7/0.3 & 2/1.5 \\
\hline
Focus residual (nm rms) & 37/52.9 & 38.4/53.1 & 35.9/53.2 & 50.5/50.8 \\
\hline
\end{tabular}
\end{center}
\end{table}

\section{CONCLUSION}

We have studied the performance of a near-infrared pyramid for the next generation AO of Keck II, which will include a classical AO mode (NGS mode) and a LGS mode. In NGS mode, the pyramid will provide a Strehl ratio in K band of  ~37\% at magnitude H = 14 and ~80\%  at high flux (20$\times$20 configuration). The latter can be increased with a finer pupil sampling (32$\times$32, or 40$\times$40) and a higher degree of correction (i. e. 32$\times$32 DM), without degrading significantly the performance at low flux. The 40$\times$40 sampling would also provide more robustness to errors such as misregistration. In LGS mode, the pyramid would not benefit from a hardware rebin of pixels, and a focal plane sensor would be preferable to estimate low orders. We have demonstrated that through a comparison with LIFT, which provides a gain of 2 magnitudes on tip/tilt up to 15" off axis and a similar performance on focus. It was also shown that LIFT gives best results with images sampled at Nyquist (15 mas pixels). In future works, we will explore more off-axis distances and seeing conditions to confirm the advantage of using LIFT. We will also study the impact of the atmosphere dispersion on both sensors.

\acknowledgments
This work was partly funded by INAF (Research Grant DD 27). The Keck II pyramid wavefront sensor is funded by the National Science Foundation under Grant No. AST-1611623.

\bibliography{pyr_sims_ao4elt5} 

\end{document}